# MeTime: An R package for reproducible longitudinal metabolomics data analysis.

Bharadwaj Marella*[,1], Patrick Weinisch*[,1], Lara Vehovec[1], Vinh Tran[1], Josef J Bless[1], Yacoub A. Njipouombe Nsangou[1], Gabi Kastenmüller[1], Matthias Arnold[#,1,2]

1. Institute of Computational Biology, Helmholtz Zentrum München, German Research Center for Environmental Health, Neuherberg, Germany.
2. Department of Psychiatry and Behavioral Sciences, Duke Institute for Brain Sciences, Department of Medicine, Duke University, Durham, NC, USA.

* Equal contribution

\# Correspondence to: matthias.arnold@helmholtz-munich.de

# Abstract

MeTime is an opensource R package for reproducible analysis of longitudinal metabolomics data. It builds upon a central S4 container, metime_analyser, that stores multiple datasets, associated metadata and analysis outputs, enabling unified handling of complex longitudinal studies. Analyses are constructed by piping modular functions, beginning with data transformations (mod_*), followed by calculations (calc_*), and optional meta-analysis (meta_*), so entire workflows remain transparent and easy to modify. MeTime wraps numerous existing methods within a consistent interface, including sample and metabolite distributions, correlation/distance matrices, dimensionality reduction (PCA, UMAP, t-SNE), random forest imputation and feature selection via Boruta, eigenmetabolites and WGCNA-based clustering, conservation index analysis, regression models (linear, mixed-effects, and generalized additive), and partial-correlation networks. By retaining all intermediate results and provenance within the container, MeTime facilitates iterative exploration and ensures reproducible reporting via automatically generated HTML/PDF outputs. Comprehensive user guides, case studies and reference documentation accompany the package, making MeTime a versatile platform for longitudinal omics workflows.

# 1. Introduction

Longitudinal metabolomics provides a powerful framework for characterizing temporal changes in metabolite profiles within an organism. By capturing within-subject dynamics over time, longitudinal designs enable the investigation of molecular responses to stimuli, interventions, and disease progression, while accounting for inter-individual variability. Metabolomic profiles provide a downstream readout of biochemical variation shaped by molecular, environmental, lifestyle, and disease-related influences. Longitudinal measurements can capture dynamic changes associated with disease progression, intervention response, and predict future health outcomes, including all-cause mortality. (Rosato et al., 2017; Castelli et al., 2021; Shen et al., 2023; Arnold et al., 2020; Lacruz et al., 2018).

Advances in mass spectrometry technologies and data acquisition workflows have led to a rapid increase in the generation of longitudinal metabolomics datasets. This growth has intensified the demand for scalable, flexible, and reproducible computational approaches capable of handling high-dimensional, multi-timepoint data (Wishart et al., 2022; Trifonova et al., 2023). The analysis of such data commonly relies on a combination of multivariate methods, regression-based approaches (Gibbons et al., 2010), conservation index analysis (Yousri et al., 2014), and network-based models to capture temporal structure and inter-metabolite relationships (Krumsiek et al., 2011; Epskamp et al., 2018).

In parallel, the metabolomics community has placed increasing emphasis on reproducibility, transparency, and open science practices (Considine et al., 2018; Du et al., 2022). Contemporary publication standards often require the sharing of source code, parameters, and intermediate results alongside manuscripts (Baker, 2016). While these practices improve scientific rigor, they also impose a substantial burden on researchers, who must design analysis pipelines that are both flexible and fully traceable.

Several toolboxes have been developed to address these challenges, including MetaboAnalyst (Chong et al., 2018), maplet (Chetnik et al., 2022), and MeTEor (Grabert et al., 2024). al, 2024), which facilitate structured workflow development, standardized reporting, and to better understand longitudinal data structures However, an integrated R package specifically designed to support the end-to-end analysis of longitudinal metabolomics data across multiple datasets, while maintaining complete pipeline transparency, remains lacking.

Here, we present MeTime, an open-source R package (https://github.com/compneurobio/MeTime/) that brings together longitudinal metabolomics analyses in a single unified framework: it stores multiple datasets in a central S4 object, wraps a range of existing analysis methods behind a consistent interface, supports modular pipeline construction, and generates self-contained reports for transparent and reproducible workflows and results. In addition to describing the software design and functionality, we demonstrate the use of MeTime through an executable longitudinal metabolomics case study, illustrating how complex analytical pipelines and their results can be generated, stored, and reported in a fully reproducible manner.

# 2. Implementation and Functionality

MeTime enables transparent and reproducible analytical pipelines for longitudinal omics data by centering all inputs, outputs, and provenance in a single S4 container, the metime_analyser. This object stores raw and processed data, analytical results, and detailed metadata describing each pipeline step (called functions, parameter settings, and execution order). Pipelines are assembled by sequentially passing the metime_analyser through S4 methods using the native R pipe operator (|>) or %>% from magrittr (Bache et al., 2022), making analyses straightforward to review, reproduce, and retrace.

A key design goal is native support for longitudinal metabolomics projects that include multiple datasets (e.g., cohorts, experimental batches, or analytical platforms) that may differ in sample size and missingness. While existing container classes such as SummarizedExperiment (Morgan et al., 2023) and data structures implemented in maplet (Chetnik et al., 2022) provide robust solutions for single datasets, they do not directly address coordinated workflows spanning multiple heterogeneous datasets. MeTime addresses this by enabling independent yet synchronized analyses across datasets while storing all derived outputs within one metime_analyser object (**Figure 1**). Meta-analysis summaries are stored separately in a dedicated meta_results class, which cleanly separates primary results from downstream meta-analysis. See **Supplementary Information** for more details on the architecture of these classes.

MeTime follows a modular philosophy in which each analytical component (e.g., feature selection, regressions, data driven networks) is implemented as an independent function that can be combined into complete workflows. Functions follow a standardized tagging scheme reflecting their role in the pipeline (**Figure 1**), enabling users to insert, remove, or reorder steps with minimal effort and support rapid prototyping. The package currently comprises ~80 functions spanning data loading, extraction (get_*), modification (mod_*), augmentation (add_*), statistical analysis (calc_*), meta-analysis of summary statistics (meta_*), and export (write_*). Statistical methods include missing value imputation, dimensionality reduction, linear models, linear mixed-effects models, generalized additive mixed models, conservation index analysis, and data driven networks, alongside utilities for publication-ready visualizations and structured result tables. Although developed for metabolomics, the underlying architecture is extensible to other omics modalities. Templates and extensive documentation are provided to facilitate extension with custom functions while preserving compatibility with the core architecture (https://github.com/compneurobio/MeTime/).

Following pipeline execution, results remain embedded in the metime_analyser and can be summarized with write_report(), which generates comprehensive reports via knitr and Rmarkdown packages (Allaire et al., 2023). Reports enumerate pipeline steps in the order executed, including function names, parameter settings, and associated tables/figures (**Figure 1**), and can be exported in PDF or HTML format for sharing and archival. By consolidating results with code provenance and parameterization in a single document, MeTime supports transparent reporting and reproducibility.

# 3. Exemplary analysis

To demonstrate the functionality of MeTime, we provide three complementary usage examples. First, the package includes the publicly available HuMet dataset (Weinisch et al., 2024), which is used in the GitHub tutorials and example workflows as a small and accessible dataset for learning the package and exploring its main functions. Second, we provide a fully reproducible workflow on GitHub (https://github.com/compneurobio/MeTime/tree/main/docs/case-studies) that replicates the association analyses using the longitudinal metabolomics dataset from the Alzheimer's Disease Neuroimaging Initiative measured with the Biocrates MXP® Quant 500 platform from Marella et al., 2025. After quality control and construction of the metime_analyser object, this analysis can be rerun directly using the provided code. Third, we use the same longitudinal dataset to show other types of analyses that can be carried out with MeTime, including dimensionality reduction, trajectory modeling with linear models, mixed-effects models, and generalized additive mixed models, conservation index analysis, and Gaussian graphical modeling (see **Supplementary Information**).

These additional analyses are provided in the **Supplementary Data** as paired script and report files, so that each workflow can be viewed both as code and as its resulting output. **Supplementary Information** gives an overview of all results along with more details about the package architecture. Together, these examples show that MeTime can be used both as an easy starting point for new users and as a reproducible framework for a wide range of longitudinal omics analyses.

# 4. Conclusion

MeTime provides a comprehensive framework for the reproducible analysis of longitudinal metabolomics data by enabling rapid construction of transparent and modular analytical pipelines. By integrating data management, statistical modeling, result storage, and report generation within a single S4-based architecture, MeTime thereby reduces the technical overhead associated with longitudinal data analysis. The executable case study presented here demonstrates how MeTime can be used to generate fully reproducible results directly from package vignettes, supporting both methodological development and transparent scientific reporting.

# 6. Author contributions

**Conceptualization & Methodology:** GK, MA; **Code development & software:** BM, PW, MA; **Funding acquisition:** GK, MA; **Testing:** BM, PW, VT, LV, YN, MA; **Writing – original draft:** BM, JJB, MA; **Writing – review & editing:** All authors.

# 7. Acknowledgements

This work was supported by the National Institutes of Health/the National Institute on Aging through grants 1RF1AG057452, R01AG069901, U01AG061359, and R01AG081322. This work was also supported by the Deutsche Forschungsgemeinschaft (DFG, German Research Foundation) FOR 5795 (HyperMet) and by the German Federal Ministry of Education and Research (BMBF) (BiomarKid, 01EA2203B and PlantIntake, 1EA2204B) under the umbrella of the European Joint Programming Initiative "A Healthy Diet for a Healthy Life" (JPI HDHL) and the ERA-NET Cofund ERA-HDHL (GA N° 696295) of the EU Horizon 2020 Research and Innovation Programme).

# 8. Conflicts of interest

All authors declared no conflicts.

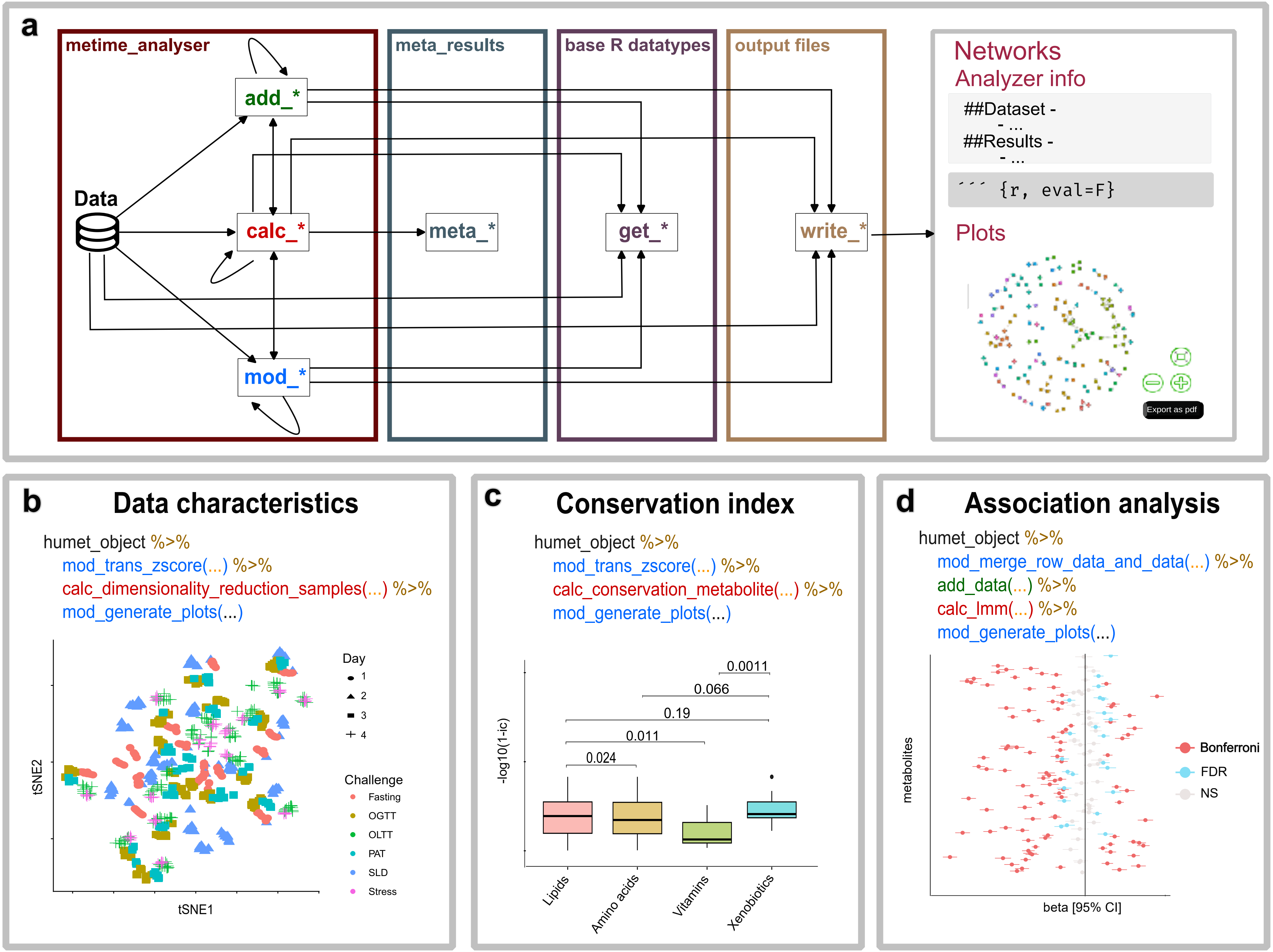

a

metime_analyser
meta_results
base R datatypes
output files

Data
add_*
calc_*
mod_*
meta_*
get_*
write_*

Networks
Analyzer info
##Dataset -
- ...
##Results -
- ...

```
``` {r, eval=F}
```

Plots
Export as pdf

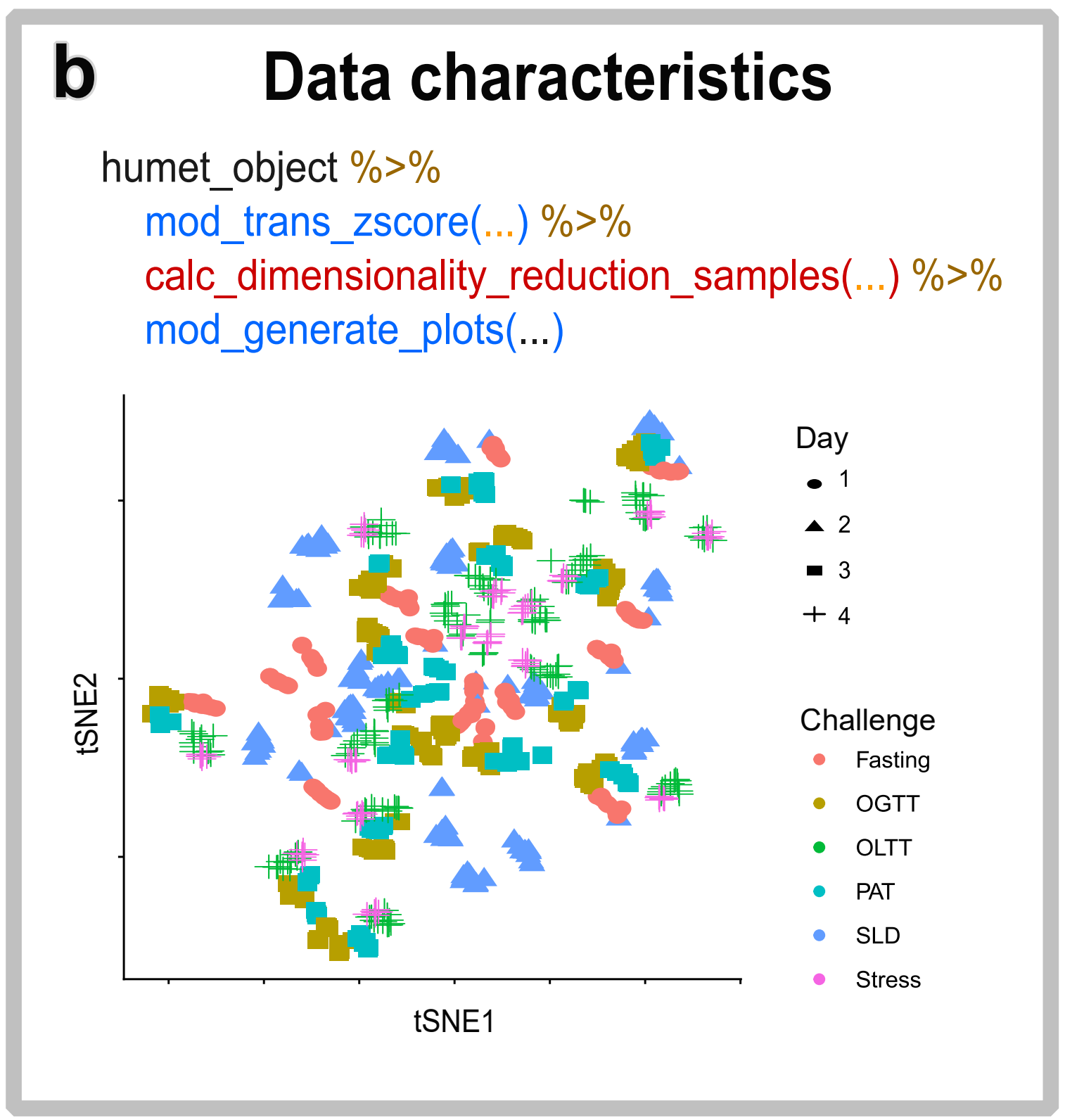


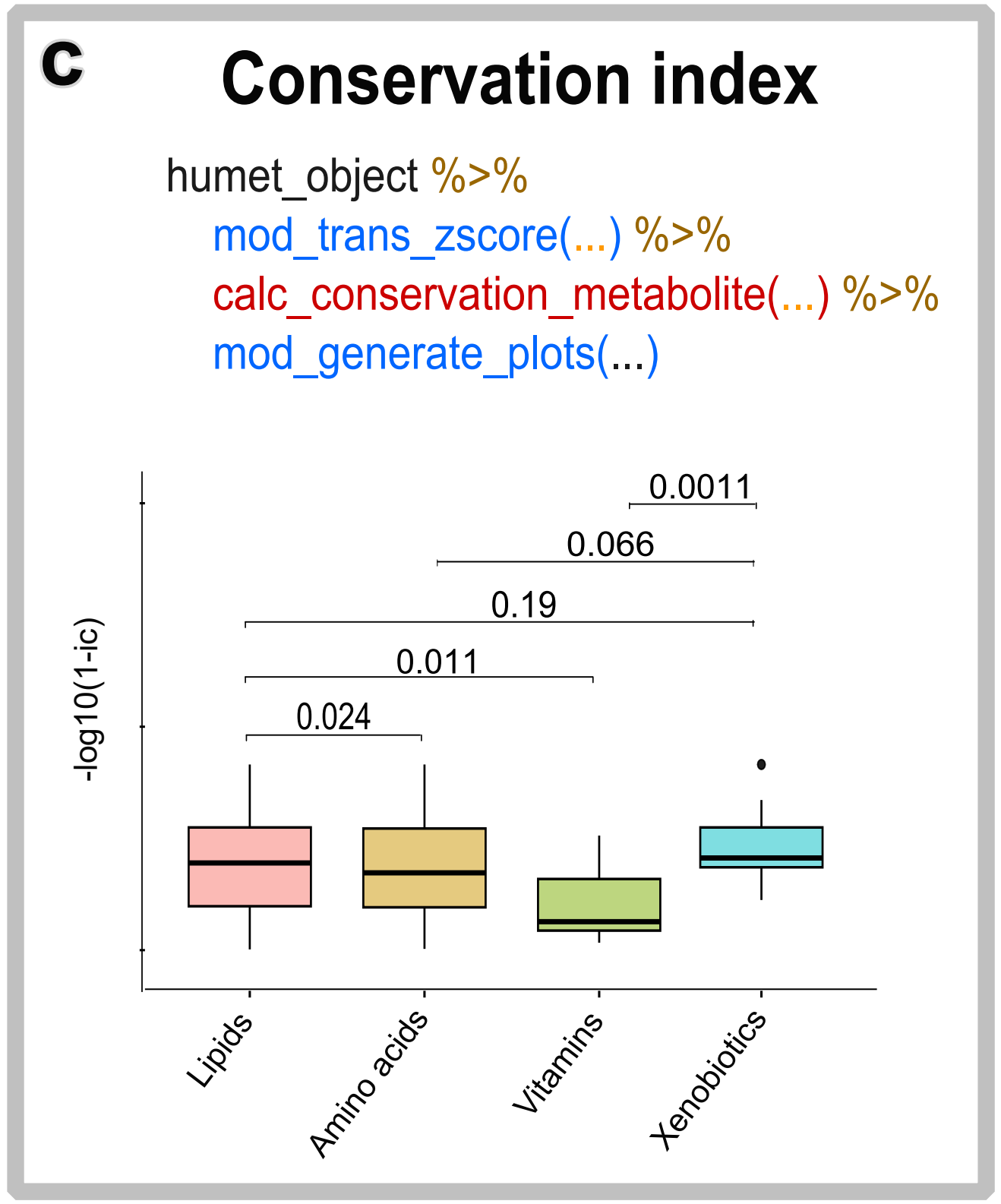


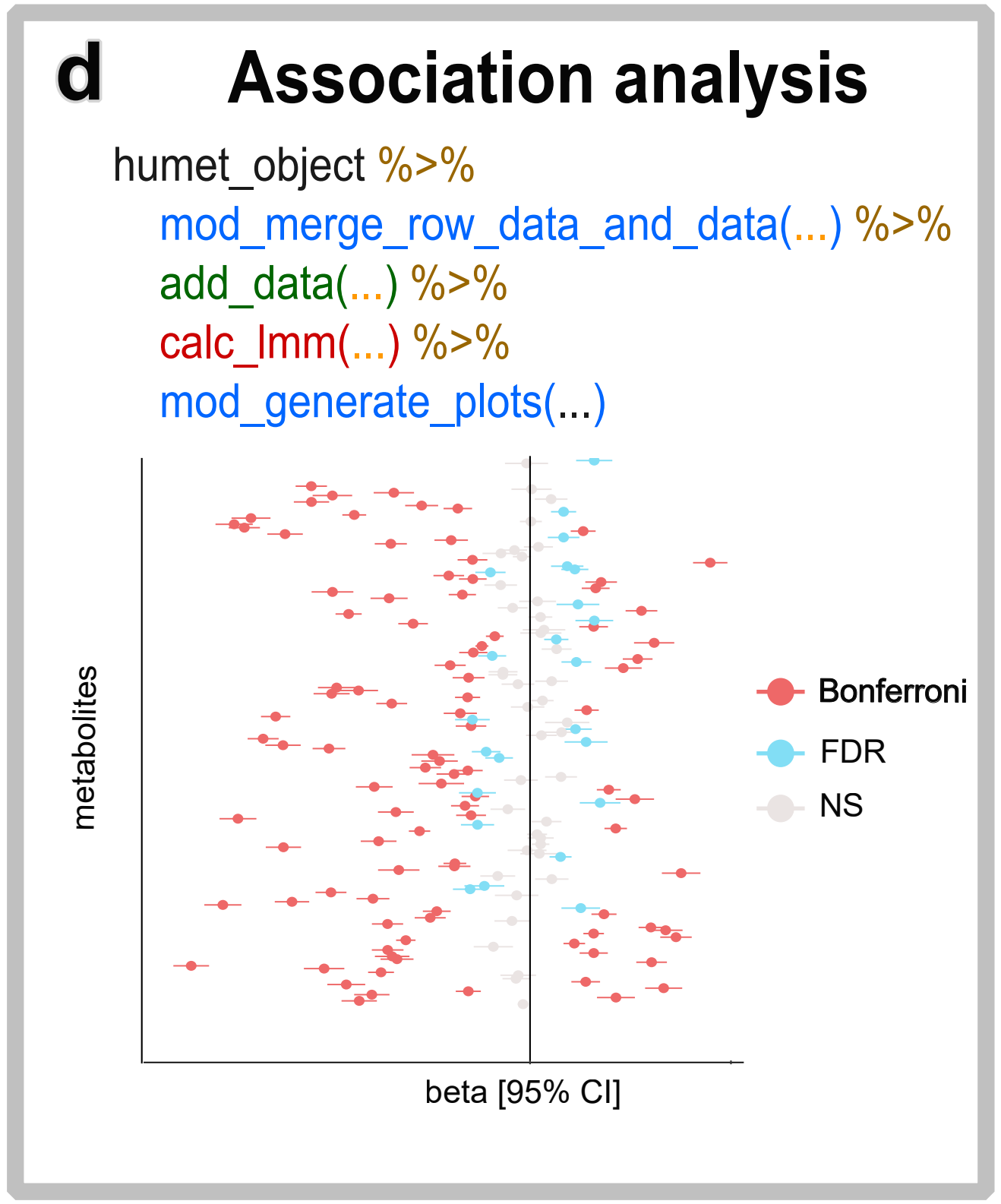
```

**Figure 1: Overview of MeTime's architecture and example analyses. (a)** A schematic of the modular workflow: the central metime_analyser object holds input data and results; functions prefixed as add_* import or merge data, mod_* modify and annotate, calc_* perform analyses, meta_* compare summary statistics, get_* retrieves contents, and write_* generates reports. Summary statistics across two calculations can be summarised in secondary meta_results objects and visualised via built-in plotting mechanisms. **(b)** Data characteristics: an example pipeline that z-scores metabolite abundances, computes a t-SNE embedding, and plots samples by challenge and day of challenge. **(c)** Conservation index: box plots showing stability of metabolic profiles across lipid and non-lipid classes after transformation and conservation index calculation. **(d)** Association analysis: effect-size estimates and 95 % confidence intervals for selected metabolites from a linear mixed-effects model; colors indicate significance after Bonferroni, FDR adjustment ($q<0.05$) and nominal ($p<0.05$) thresholds.